\documentstyle[12pt]{article}

\newcommand{\be}{\begin{equation}}
\newcommand{\ee}{\end{equation}}
\newcommand{\bea}{\begin{eqnarray}}
\newcommand{\eea}{\end{eqnarray}}
\newcommand{\nn}{\nonumber \\}

\newcommand{\ba}{\begin{array}}
\newcommand{\ea}{\end{array}}
\newcommand{\vs}[1]{\vspace{#1 mm}}

\def\bbox{{\,\lower0.9pt\vbox{\hrule \hbox{\vrule height 0.2 cm
\hskip 0.2 cm \vrule height 0.2 cm}\hrule}\,}}
\newcommand{\dsl}{\pa \kern-0.5em /}

\newcommand{\pa}{\partial}

\font\mybb=msbm10 at 10pt
\def\bb#1{\hbox{\mybb#1}}

\def\bE {\bb{E}}

\begin{document}

\topmargin 0pt
\oddsidemargin 5mm

\renewcommand{\thefootnote}{\fnsymbol{footnote}}
\begin{titlepage}

\begin{center}
{\Large The M(atrix) model/$adS_2$ correspondence\footnote{To appear in
proceedings of the Third Puri Workshop on Quantum Field Theory,
Quantum Gravity and Strings, Puri, Orissa, India, 9-19 December 1998.}}
\vs{10}

{\large P.K. Townsend} \\
\vs{5}
{\em DAMTP, University of Cambridge,\\ 
Silver Street, Cambridge CB3 9EW, UK.}
\end{center}
\vs{10}
\centerline{{\bf Abstract}}

The M(atrix) model has a dual realization as IIA superstring theory
in the near-horizon geometry of the supergravity D0-brane. The role of $adS_2$
in this correspondence is reviewed and some aspects of holography that it
suggests are discussed. 

\end{titlepage}
\newpage 
\renewcommand{\thefootnote}{\arabic{footnote}}
\setcounter{footnote}{0}

The dynamics of $n$ D0-branes of IIA superstring theory at an energy scale $E$ 
is described, in the limit
\be
\sqrt{\alpha'}E\rightarrow 0\, , \qquad g_s \rightarrow 0\, ,
\ee
by a {\it non-relativistic} $U(n)$ supersymmetric gauge quantum mechanics
with 16 supersymmetries, otherwise known as the M(atrix) model \cite{BFSS}.
Here, $\alpha'$ is the inverse string tension and $g_s$ is the string coupling
constant. The M(atrix) model is just a D=10 super-Yang-Mills (SYM) theory
dimensionally reduced to D=1. The (dimensionful) coupling constant of this 
SYM theory is
\be
g_{YM} = g_s^{1/2}(\alpha')^{-3/4}\, .
\ee

The group of symmetries of the M(atrix) model (excluding supersymmetries) is
the D=10 Bargmann group, which is a central extension  of the Galilean group;
the central charge is the D0-brane mass. This group is a subgroup of the D=11
Poincar{\'e} group for which a null component of the 11-momentum is central.
The (super)Bargmann invariance of the $U(1)$ theory follows from the fact that
the action is the null reduction of the action for the D=11 massless
superparticle \cite{bergtown}; the extension to $U(n)$ then follows from the 
fact that the relative D0-brane coordinates of translation 
and boost invariant. We
conclude that the non-relativistic limit described above is equivalent to a
limit in which the spacelike circle of $S^1$ compactified M-theory becomes
lightlike. It has been argued that all degrees of freedom of IIA superstring
theory other than D0-branes decouple in this limit \cite{seiberg}, so that the
M(atrix) model provides a definition of M-theory on a lightlike circle, as
originally conjectured \cite{BFSS,lenny}. 

According to the M(atrix) model, the UV regime of D=11 supergravity is
described by the IR dynamics of the SYM theory. But the IR limit of the
gauge theory is its strong coupling limit. This can be investigated in
the limit of large $n$ by
't Hooft's topological expansion \cite{thooft}, for which the effective
dimensionless coupling constant at the energy scale $E$ is
\be
g_{eff}(E) = {g_{YM} N^{1/2}\over E^{3/2}}\, .
\ee
The topological expansion is an asymptotic expansion in $g_{eff}$. According to
a generalization of the $adS/CFT$ correspondence \cite{IMSY,BST}, the dual
asymptotic expansion in $g_{eff}^{-1}$ is provided by IIA supergravity in the
background of its D0-brane solution \cite{HS}. In the {\it string frame} the
non-vanishing fields of this solution are
\bea\label{stringframe}
ds_{st}^2 &=& -H^{-{1\over2}}dt^2 + H^{1\over2}ds^2(\bE^9) \nn
e^\phi &=& g_s H^{3\over4} \nn
\tilde F_8 &=& g_s^{-1}\star_9 dH
\eea
where $\tilde F_8$ is the 8-form dual of the 2-form Ramond-Ramond (RR) field
strength, $\star_9$ is the Hodge dual on $\bE^9$, and $H$ is a harmonic
function on $\bE^9$. The $g_s$-dependence may be determined from the solution
with $g_s=1$ by means of the transformation 
\bea
\phi &\rightarrow& \phi + \lambda\nn
\tilde F_8 &\rightarrow &e^{-\lambda}\tilde F_8
\eea
where $\lambda$ is a constant. This is not an 
invariance of the action but it is
an invariance of the field equations.

In coordinates such that
\be
ds^2(\bE^9) = dr^2 + r^2d\Omega_8^2\, ,
\ee
where $d\Omega_8^2$ is the $SO(9)$-invariant metric on $S^8$, we may choose 
the harmonic function $H$ to be
\be
H= 1+ g_sN\left({\sqrt{\alpha'}\over r}\right)^7\, .
\ee
Given the factor of $g_s^{-1}$ in $\tilde F_8$, this choice corresponds to N
coincident D0-branes at the origin of $\bE^9$. We can now rewrite $H$ as
\be
H = 1+  {g^2_{eff}(U) \over (\sqrt{\alpha'}U)^4}
\ee
where
\be
U= r/\alpha'\, .
\ee
The variable $U$ has dimensions of energy. It is the energy of a string of 
length $r$, although one should not read too much into this fact as $U$ will
shortly be seen to be merely a convenient intermediate variable. For the moment
we need only suppose (subject to later verification) that `low energy'
corresponds to the limit $\sqrt{\alpha'}U\rightarrow 0$. For any non-zero
$g_{YM}$ this implies $g_{eff}(U)\rightarrow\infty$, which we need in any case
for the validity of the dual IIA supergravity description of the D0-brane
dynamics. The low-energy limit is therefore a `near-horizon' limit in the sense
that
\be
H \rightarrow {g^2_{eff}(U) \over (\sqrt{\alpha'}U)^4}\, .
\ee
There is a problem with this limit, however, because the string frame metric of
(\ref{stringframe}) has a curvature singularity at singularities of $H$,
i.e. at $U=0$. 

Although the string frame is natural in the context of IIA superstring
theory, it is not obviously the preferred frame in the context of the M(atrix)
model. Of course, no frame is really `preferred' because the physics cannot 
depend on the choice of frame, but there may be a frame in which the physics
is simplest. It was argued in \cite{BST} that the preferred frame in
this sense is the `dual' frame, defined for a general p-brane (up to homothety)
as the one for which the dual brane (the D6-brane in our case) has a tension
independent of the dilaton. In this frame, and for all $p\ne 5$,  the
singularities of the harmonic function $H$ in the p-brane metric are Killing
horizons near which the D-dimensional geometry is $adS_{p+2}\times S^{D-p-2}$
\cite{DGT,BPS}. This  result generalizes the 
interpolation property of p-branes,
such as the M2,M5 and D3 branes, that do not couple to a dilaton \cite{GT}. For
the D0-brane the dual frame metric $d\tilde s^2$ is related to the string frame
metric as follows:
\be
d\tilde s^2 = (e^\phi N)^{-{2\over7}} ds^2_{st}\, .
\ee
The factor of $N$ is included here for later convenience. The D0-brane metric 
is now
\be
d\tilde s^2 = \left(g_s N\right)^{-{2\over7}}\left[-H^{-{5\over7}}dt^2
+ H^{2\over7}(dr^2 + r^2d\Omega_8^2)\right]\, ,
\ee
and in the near-horizon limit we have
\be
d\tilde s^2 \sim \alpha'\left[ -\left(g_{YM}^2N\right)^{-1} U^5 dt^2
+ U^{-2} dU^2 + d\Omega_8^2 \right]\, .
\ee
The singularity of the metric at at $U=0$ is now only a coordinate singularity
at a Killing horizon of $\partial_U$, but the metric still depends on
the SYM coupling constant. To circumvent this, 
we define the a new radial variable $u$ (with dimensions of energy) by
\be
u^2 = {U^5\over g^2_{YM} N}\, .
\ee
For convenience we also introduce the rescaled time coordinate
\be
\tilde t= {5\over 2}\, t\, .
\ee
The near-horizon D0-brane solution is now \cite{BST}
\bea\label{holog}
(\alpha')^{-1} d\tilde s^2 &=& {4\over 25}\left[-u^2
d\tilde t^2 + u^{-2} du^2\right] + d\Omega_8^2\nn
e^\phi &=& N^{-1} \left[g_{eff}(u)\right]^{7/5}\nn
\tilde F_8 &=& 7N\, vol(S^8)
\eea
We recognise this as $adS_2\times S^8$, with standard (horospherical)
coordinates for the $adS_2$ factor. As $u$ is now the 
only dimensionful variable
it sets the energy scale for solutions of the massless wave equation in the
near-horizon geometry. This fact implies that an infra-red cut-off of
supergravity at length $\alpha' u$ corresponds, via holography \cite{WS}, to an
ultraviolet cut-off of the D0-brane SYM theory at energy $u$ \cite{PP}. 

The $adS_2$ metric has an $SL(2;R)$ isometry group. This does not extend to a
symmetry of the full near-horizon solution because the dilaton field is 
invariant only under the one-dimensional subgroup generated by $\partial_t$,
However, scale transformations generated by the Killing vector field
$t\partial_t -u\partial_u$ take one hypersurface of constant $u$ into another
such hypersurface, which leads to a rescaling of $g_{eff}(u)$. A hypersurface
of constant $u$ is thus the vacuum of the M(atrix) model at coupling
$g_{eff}(u)$. As we rescale $u$ we go either to a free theory with $g_{eff}=0$
at the $adS_2$ boundary, which is obviously scale invariant, or towards a
strongly coupled theory at the Killing horizon of $\partial_t$ at $u=0$.
In order to keep $e^\phi$ small in the latter limit we must take $N$ large.
However, for any finite $N$ the effective string coupling constant will still
become large sufficiently near $u=0$ and the IIA supergravity description will
break down. This is an indication that we should pass to D=11
supergravity.

Given that the full D0-brane solution (\ref{stringframe}) is the reduction of
the M-wave, one might wonder what the near-horizon limit of the D0-brane
solution lifts to in D=11. In view of the fact that the non-relativistic
D0-brane action is the null reduction of the D=11 massless 
superparticle action,
the obvious guess is that the near-horizon limit of the D0-brane solution is a
null reduction of the M-wave. This is true, in the following sense \cite{BBPT}.
The M-wave metric is
\be\label{mwave}
ds^2_{11} = dudv + K du^2 + ds^2(\bE^9)
\ee
where $K$ is harmonic on $\bE^9$; it is also an arbitrary function
of $u$ but in order to reduce to D=10 we must choose it to be $u$-independent.
The choice $K = Q/r^7$ where $r$ is the distance from the origin in $\bE^9$
now leads to the D0-brane solution (\ref{stringframe}) after reduction along
orbits of the timelike Killing vector field $\partial_u-\partial_v$. This is
the standard timelike reduction. We may instead reduce along orbits of the
Killing vector field $\partial_u$, which is null at spatial infinity. To this
end we set $v=2t$ and write (\ref{mwave}) as
\be
ds^2_{11} = K(du + K^{-1}dt)^2 + K^{-1/2}ds^2_{st}
\ee
where 
\be
ds^2_{st} = -K^{-{1\over2}}dt^2 + K^{1\over2}ds^2(\bE^9)\, .
\ee
is the string frame 10-metric. The IIA solution resulting from reduction on
orbits of $\partial_u$ is therefore the D0-brane solution in the near-horizon
limit, i.e. with $H$ replaced by $K=H-1$. 

It is satisfying that the dual supergravity description of $n$ D0-branes for
large $n$ is a wave solution because this is what one would expect 
from the Bohr correspondence principle.  However, we have 
still to consider what happens at
$u=0$. In many cases, singularities of IIA solutions are resolved by their D=11
interpretation \cite{revisited}, but this does not 
happen here. The singularities
of the harmonic function $K$ are curvature 
singularities of the M-wave solution,
so the D=11 supergravity description must break down there. The reason that the
IIA supergravity dual description  breaks down is that the effective string
coupling becomes large. While this implies a decompactification to D=11 it also
implies that the neglect of string loop corrections, and hence M-theory
corrections in D=11, cannot be ignored. These are UV corrections
to D=11 supergravity that should be determined by the IR physics of
the M(atrix) model, but this is its strong coupling limit that
we hoped to understand via its supergravity dual. 

Although we have failed to learn much about the IR physics of the M(atrix) 
model from its supergravity dual, we can presumably learn how to
resolve the singularity of the M-wave solution of D=11 supergravity
from the IR physics of the M(atrix) model; it is just that the
M(atrix) model/$adS_2$ correspondence doesn't help us to accomplish
this. On the positive side, it does shows that the M(atrix) model 
proposal is a close cousin of Maldacena's adS/CFT proposal
\cite{malda} (as argued by other means in \cite{mart,chep}).
The latter is generally considered to provide an illustration of
the concept of holography \cite{holog}. If this is  extended to the M(atrix)
model \cite{suss} and, more generally, to other branes then the general 
statement would seem to be that the bulk gravitational physics is determined by
the physics of the `matter' on branes. M-theory provides a natural
realization of this idea (which is also suggested by the global nature of
observables in general relativistic theories) because the uniqueness
of D=11 supergravity ensures the absence of bulk matter. 

This is all rather similar to Mach's principle, as Ho{\v r}ava has previously
pointed out in the context of an alternative proposal 
for the degrees of freedom
of M-theory \cite{horava}.  The utility of 
Mach's principle is rather limited for
asymptotically flat spacetimes because the local inertial frames are then
predominantly determined by the existence of asymptopia. 
Holography is similarly
limited; its applicability to adS spacetimes is 
evidently linked to the fact that
timelike spatial infinity can be interpreted as a brane. This limitation
would not be a problem if the universe were spatially closed, but this invokes
cosmology to resolve an apparently unrelated problem. Perhaps this is an
indication that they are not unrelated and that a consistent nonperturbative
formulation of quantum gravity must incorporate cosmology. 
 
\vskip 0.5cm
{\bf Acknowledgements}:
This article is largely an elaboration for $p=0$ of work reported for general
$p$ in \cite{BST}. I thank my collaborators HarmJan Boonstra and Kostas
Skenderis, and also Gary Gibbons and Arkady Tseytlin for discussions on related
issues. I thank Mohab Abou-Zeid, Melanie Becker, Iouri Chepelev and 
Petr Ho{\v r}ava for helpful comments on an earlier version, and the 
organisers of the 3'rd Puri workshop for the opportunity to visit India.


\end{document}